\pdfoutput=1

\documentclass[aps,prb,reprint,showpacs,superscriptaddress]{revtex4-1}

\usepackage{graphicx}
\usepackage{graphics}
\usepackage{amsmath}
\usepackage{amssymb}
\usepackage{amsfonts}
\usepackage{dcolumn}
\usepackage{dsfont}
\usepackage{latexsym}
\usepackage{rotating}
\usepackage{color}
\usepackage{latexsym}
\usepackage{bbm}
\usepackage{subfigure}
\usepackage{float}
\usepackage{epsfig}
\usepackage{epsf}
\usepackage{psfrag}
\usepackage{bm}
\usepackage{amsthm}
\usepackage{eucal}
\usepackage{mathrsfs}
\usepackage{url}
\usepackage{braket}

\usepackage{color} 

\usepackage{hyperref}
\hypersetup{
colorlinks=true,final=true,
        linkcolor=blue,
        citecolor=blue,
        filecolor=blue,
        urlcolor=blue,
}
\begin{document}
\title{Role of chemical pressure on the electronic and magnetic properties of spin-1/2 kagome mineral averievite}

\author{Dibyendu Dey}
\email{ddey3@asu.edu}
\affiliation{Department of Physics, Arizona State University, Tempe, AZ - 85287, USA}

\author{Antia S. Botana}
\email{antia.botana@asu.edu}
\affiliation{Department of Physics, Arizona State University, Tempe, AZ - 85287, USA}
\date{\today}

\begin{abstract}
We investigate the electronic and magnetic properties of the kagome mineral averievite (CsCl)Cu$_5$V$_2$O$_{10}$ and its phosphate analog (CsCl)Cu$_5$P$_2$O$_{10}$ using first-principles calculations. The crystal structure of these compounds features Cu$^{2+}$ kagome layers sandwiched between  Cu$^{2+}$-P$^{5+}$/Cu$^{2+}$-V$^{5+}$ honeycomb planes,  with pyrochlore slabs made of corner-sharing Cu-tetrahedra being  formed. The induced chemical pressure effect upon substitution of V by P causes significant changes in the structure and magnetic properties. Even though the in-plane antiferromagnetic (AFM) coupling (J$_1$) within the kagome layer is similar in the two materials, the inter-plane AFM coupling (J$_2$) between kagome and honeycomb layers is five times larger in the P-variant increasing the degree of magnetic frustration in the constituting Cu-tetrahedra. 
\end{abstract}

\maketitle

\section{Introduction}

Quantum spin liquids (QSL) represent a novel state of matter in which the constituent spins are highly correlated but still fluctuate so strongly that they prevent long-range magnetic order down to zero temperature~\cite{AnderMRB, BaleNat, NormanRMP, SavaryRPP, ZhouRMP}. These characteristics make them distinct phases of matter, able to display unique exotic behavior such as new types of topological order~\cite{LawlerPRL, MazinNC}, excitations with fractional quantum numbers~\cite{HanNat}, or certain forms of superconductivity~\cite{BaskSSC, AnderSc}. 

Typically, QSLs are realized in lattices that act to frustrate the appearance of magnetism~\cite{NormanRMP, ZhouRMP}. In two dimensions, the prototypical example is the kagome lattice with spin-1/2 ions, that gets realized in a variety of minerals. Herbertsmithite (a copper hydroxy-chloride mineral with a Cu$^{2+}$: d$^9$ kagome lattice) has been intensively studied in this context~\cite{HeltonPRL, HanNat, ShoresJACS, ColeCM, JesPRB13, IqbalPRB}. This material does not show any signature of long-range magnetic order close to 0 K~\cite{HeltonPRL, HanNat}, and neutron scattering experiments exhibit a spinon continuum, an indication of fractionalized excitations~\cite{HanNat}. However, a major issue concerning herbertsmithite is its intrinsic disorder~\cite{LeeNatMat, dePRL, FreeJACS} and the fact that doping, desirable to give rise to superconductivity, is difficult to achieve~\cite{KellyPRX}. For these reasons, since the discovery of herbertsmithite, other candidate materials have been intensively searched for ~\cite{HanPRL, JesPRB, BotanaPRB, BoldPRB}.

\begin{figure}
\includegraphics[width=8.8cm]{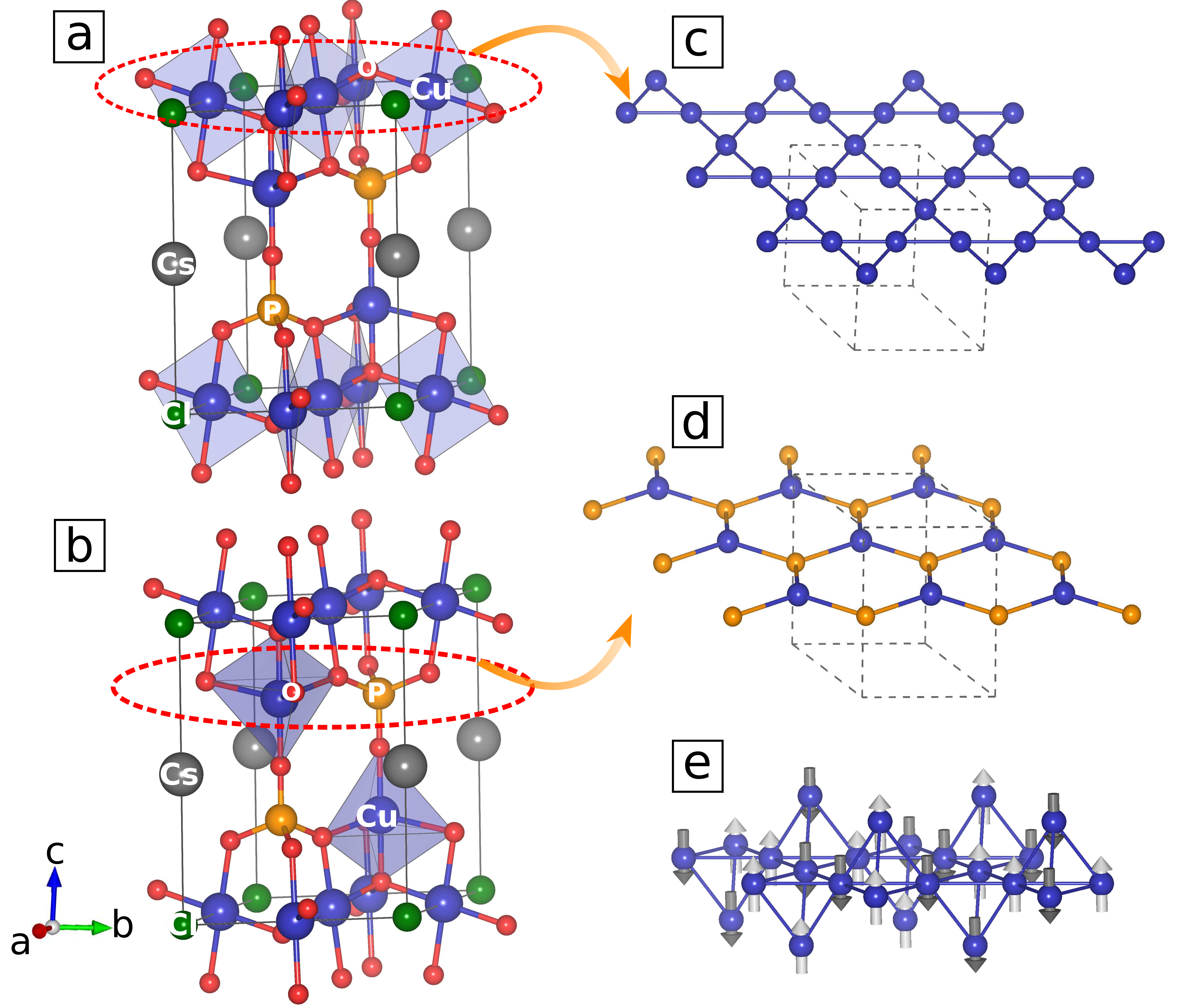}
\caption{(Color online) Crystal structure of P-averievite (V-averievite is isostructural) exhibiting square-planar (a) and trigonal bipyramidal (b)  environments for the kagome and honeycomb Cu atoms, respectively. Cu-kagome layers (c) are sandwiched between two  Cu-P-honeycomb layers (d). Each
of these trilayer blocks is separated along the $c$ axis by CsO$_2$ layers.  (e) Cu atoms of adjacent honeycomb and kagome layers form a pyrochlore slab comprised of corner-sharing Cu-tetrahedra. The corresponding AFM configuration within the Cu-tetrahedra is shown. Cs atoms are shown in gray, Cu atoms in blue, O atoms in red, P atoms in golden yellow, and Cl atoms in green.}
\label{Fig1}
\end{figure}

(CsCl)Cu$_5$V$_2$O$_{10}$ (V-averievite)~\cite{Aver_ST}, a fumarolic oxide mineral containing a spin-1/2 kagome lattice, has been synthesized and studied in the context of QSL physics~\cite{BotanaPRB}. 
The crystal structure of V-averievite contains Cu$^{2+}$ (spin-1/2)  kagome layers in which the oxygen environment of the Cu ions is square-planar. Each of these kagome planes is sandwiched between two honeycomb layers formed by Cu$^{2+}$ (with a trigonal bipyramidal environment) and V$^{5+}$ (with a tetrahedral environment) as shown in Fig. \ref{Fig1}(a-d). The Cu layers form a pyrochlore slab  comprised of corner-sharing tetrahedra. This tetrahedral geometry introduces magnetic  frustration, as interactions within the Cu$_4$ units are antiferromagnetic (AFM) as shown in Fig. \ref{Fig1}(e).  Susceptibility data show an AFM phase transition takes place at T$_N$ = 24 K in V-averievite~\cite{BotanaPRB}. In a recent experiment~\cite{WiniACS}, Winiarski {\it et al.} successfully synthesized (CsCl)Cu$_5$P$_2$O$_{10}$ (P-averievite)- the phosphate analog of V-averievite. Substitution of V by P gives rise to differences in  structural and magnetic behavior likely due to chemical pressure -the ionic radius of P$^{5+}$ in tetrahedral coordination is two times smaller than that of of V$^{5+}$. Magnetization measurements reveal strong geometric frustration with the susceptibility of P-averievite showing an AFM or spin glass-like transition at a lower temperature of 3.8 K~\cite{WiniACS}. 

In order to understand the change in magnetic response upon chemical pressure in averievite, we use first-principles calculations to obtain a microscopic magnetic model of V- and P- variants. Our results reveal important differences between the two materials, in particular, the inter-layer coupling between Cu-kagome and Cu-honeycomb atoms is five times larger in the P-variant, increasing the degree of magnetic frustration within Cu-tetrahedra that constitute a given pyrochlore slab.

\section{Computational Details}
\label{meth}
Density functional theory (DFT)~\cite{KohnShamPR, HohenKohnPR}-based calculations have been performed by using a plane-wave basis set and projector-augmented wave (PAW) potentials~\cite{BlochPAW, KressePAW}, as implemented in the Vienna {\it ab initio} simulation package VASP~\cite{KresseVASP1, KresseVASP2}. The wave functions were expanded in the plane-wave basis with a kinetic energy cutoff of 500 eV, and the reciprocal space integration was carried out with a $\Gamma$ centered k-mesh of 8$\times$8$\times$6. For the exchange-correlation functional, the Perdew-Burke-Ernzerhof (PBE)~\cite{GGAPBE} version of the generalized gradient approximation (GGA) has been used for non-magnetic calculations. For undoped V- and P-averievite, experimental structural parameters as obtained from synchrotron X-ray diffraction data~\cite{BotanaPRB, WiniACS} have been used in our calculations. For Zn-doped P-averievite, a full structural relaxation has been performed within GGA in the non-magnetic state until the resulting forces become significantly small (0.01 eV/\AA). In order to construct different magnetic configurations, a 2$\times$2$\times$1 supercell containing four formula units has been used for both undoped and doped compounds. In spin-polarized calculations, strong correlation effects for the Cu 3$d$-electrons  have been incorporated within GGA+U using the Dudarev approach~\cite{DudarevPRB}. We have used an effective on-site Coulomb repulsion~\cite{Hubbard} U$_{eff}$= U-J = 5 eV, reasonable for this type of Cu-material~\cite{BotanaPRB}.
Hopping integrals have been obtained from maximally localized Wannier functions (MLWF), constructed using WANNIER90~\cite{PizziWAN}.
\begin{figure}
\includegraphics[width=8.4cm]{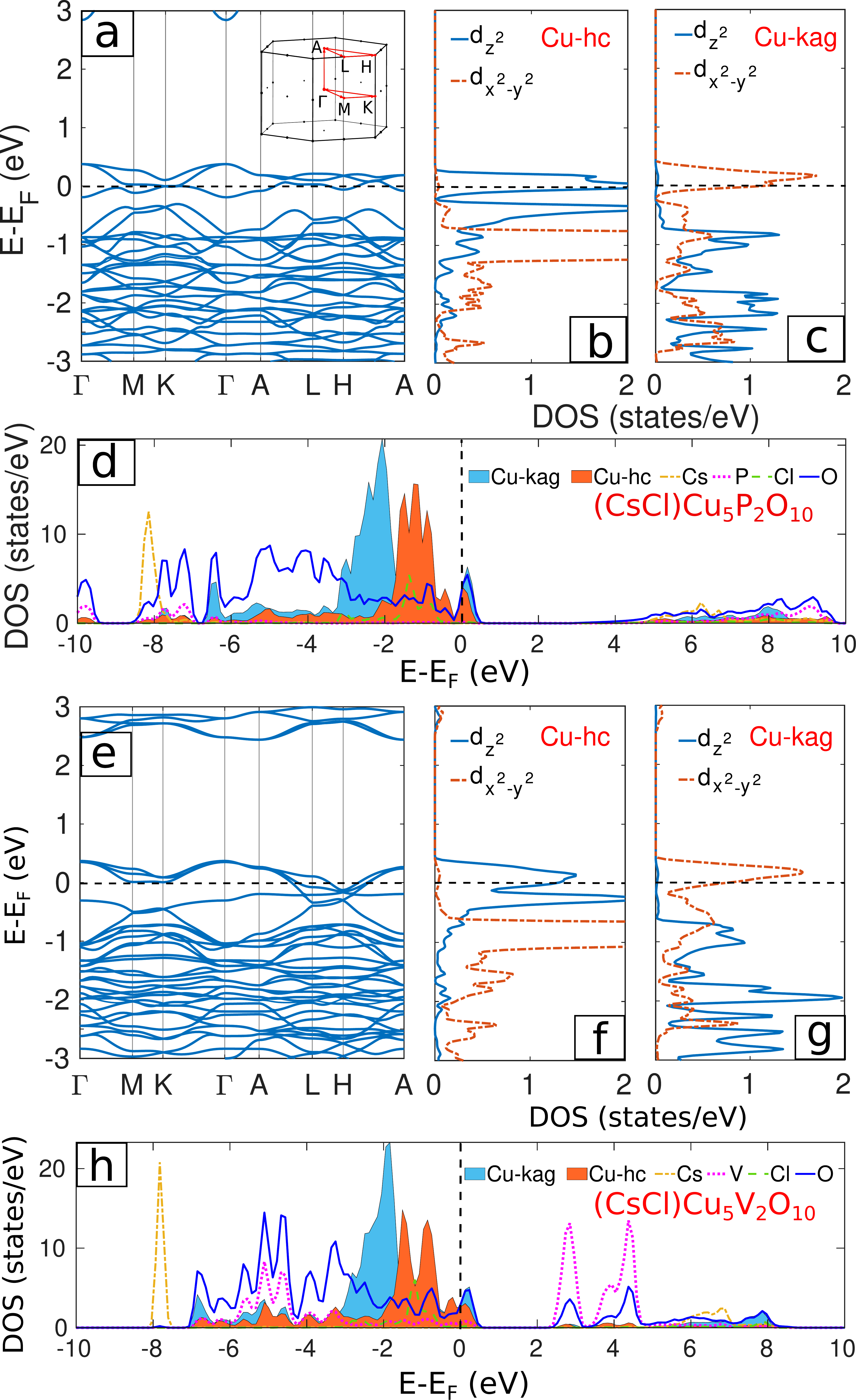}
\caption{\label{Fig2}(Color online) Non-magnetic electronic structure of P-averievite (top panels, a-d) and V-averievite (bottom panels, e-h) within GGA. (a,e) Band structure, Cu-$d_{x^2-y^2}$ and $d_{z^2}$ DOS for  honeycomb (b,f), and kagome (c,g) Cu atoms, (d,h) atom projected DOS. The band structures are shown  along the high-symmetry points $\Gamma$=(0,0,0), M=(1/2,0,0), K=(1/3,1/3,0), A=(0,0,1/2), L=(1/2,0,1/2), and H=(1/3,1/3,1/2) of the Brillouin zone (BZ) (shown as an  inset of panel (a)).}
\end{figure}

\begin{table}
\centering
\begin{tabular}{p{3cm} p{2.5cm} p{2.5cm} p{0.1cm}}
\hline
 \hline
\centering   & \centering (CsCl)Cu$_5$P$_2$O$_{10}$ & \centering (CsCl)Cu$_5$V$_2$O$_{10}$ & \\
\hline
\centering Cu$_k$-Cu$_k$ & \centering 3.09 & \centering 3.18 &\\ 
\centering Cu$_k$-O$_k$ & \centering 1.87 & \centering 1.88 &\\ 
\centering Cu$_k$-O$_h$ & \centering 1.98 & \centering 2.03 &\\ 
\centering Cu$_h$-O$_h$ & \centering 2.15 & \centering 2.10 &\\ 
\centering Cu$_h$-O$_k$ & \centering 1.83 & \centering 1.85 &\\ 
\hline
\hline
\end{tabular}
\caption{Nearest-neighbor Cu-Cu distances in the kagome plane, and Cu-O distances for P-averievite and V-averievite from experimental structural data~\cite{BotanaPRB, WiniACS}. Cu$_{k}$/O$_{k}$ (Cu$_{h}$/O$_{h}$) indicate Cu/O atoms in the kagome (honeycomb) planes, respectively. All distances are in \AA.}
\label{Tab1}
\end{table}

\section{Results and Discussion}
\label{res}
\subsection{Structural and electronic properties}
\label{struct}
P-averievite crystallizes in a trigonal P$\overline{3}$m1 space group at room temperature, and below 12 K the system undergoes a transition to a monoclinic phase~\cite{WiniACS}. A similar effect has been reported in V-averievite \cite{BotanaPRB}. Since the structural details of the low-temperature phase are not available, we considered the P$\overline{3}$m1 structure in our calculations for both materials. The crystal structure of P-averievite and its constituent kagome and honeycomb planes, as well as the environments for each Cu (square-planar in the kagome planes, trigonal bipyramidal in the honeycomb ones), are shown in Fig.~\ref{Fig1}(a)-(d). The tetrahedra formed by Cu atoms of adjacent honeycomb and kagome layers and the corresponding pyrochlore slab can be seen in Fig.~\ref{Fig1}(e). Bond lengths for P- and V-averievite from experimental structural data are summarized in Table~\ref{Tab1}. The phosphate material exhibits a lower volume and shorter Cu-Cu and Cu-O bonds due to the smaller size of P- in tetrahedral coordination the ionic radii of P$^{5+}$ and  V$^{5+}$  are 0.17 and 0.36 \AA, respectively. The Cu$_k$-O$_k$-Cu$_k$ and Cu$_k$-O$_k$-Cu$_h$ bond angles are $\sim$ 100$^\circ$ in both materials- the latter being 4$^\circ$ larger in P-averievite (106$^\circ$ vs. 102$^\circ$). 

\begin{table}
\centering
\begin{tabular}{p{3cm} p{2.5cm} p{2.5cm} p{0.1cm}}
\hline
 \hline
\centering $\vert t_{pd} \vert$ (eV) & \centering (CsCl)Cu$_5$P$_2$O$_{10}$ & \centering (CsCl)Cu$_5$V$_2$O$_{10}$ & \\
\hline
\centering Cu$_{k}$-d$_{x^2-y^2}$-O$_{k}$-p$_{x,y}$ & \centering 0.55 & \centering 0.55 & \\
\centering Cu$_{k}$-d$_{x^2-y^2}$-O$_{h}$-p$_{x,y}$ & \centering 1.15 & \centering 1.04 &\\ 
\centering Cu$_{h}$-d$_{z^2}$-O$_{k}$-p$_{x,y}$ & \centering 0.97 & \centering 0.83 &\\ 
\centering Cu$_{h}$-d$_{z^2}$-O$_{h}$-p$_{x,y}$ & \centering 0.40 & \centering 0.37 &\\
\hline
\hline
\end{tabular}
\caption{Leading hopping integrals ($\vert t_{pd} \vert$) for P and V-averievite calculated from  Wannier functions. Cu$_k$(O$_k$) and Cu$_h$(O$_h$) represent the copper(oxygen) atoms in the kagome and honeycomb layers, respectively.}
\label{Tab2}
\end{table}

We now analyze the effects of P-substitution on the non-magnetic electronic structure of averievite.  Fig.~\ref{Fig2} shows the non-magnetic GGA band structures, as well as the orbital, and atom-resolved density of states (DOS) for both P- (top panels) and V-averievite (bottom panels). A metallic state is obtained  with five Cu bands in the vicinity of the Fermi level (E$_F$) in both materials (Fig. \ref{Fig2} (a), (e)) in agreement with a Cu$^{2+}$: d$^{9}$ configuration. These correspond to 3-d$_{x^2-y^2}$ kagome-Cu bands and 2-d$_{z^2}$ honeycomb-Cu bands as reflected in the orbital resolved DOS. This is a consequence of the fact that for the kagome Cu (with square-planar coordination), the d$_{x^2-y^2}$ orbital lies the highest in energy. In contrast, for the honeycomb Cu (with trigonal-bipyramidal coordination and short Cu-O apical bonds along the c axis), the d$_{z^2}$ orbital is the highest in energy. Interestingly, among the five Cu bands, two are separated from the other three bands in P-averievite, unlike in V-averievite. The total DOS plots (Fig. \ref{Fig2}(d), (h)) show the large degree of hybridization between O-2p and Cu-3d states in both systems. An important difference is that the valence bandwidth in P-averievite is $\sim$8.5 eV, increased by 1.5 eV with respect to its V-counterpart ($\sim$7 eV) due to the induced chemical pressure effect upon V by P substitution. In addition, the unoccupied V-d states in V-averievite lie approximately 2 eV above the Fermi level (consistent with a V$^{5+}$: d$^0$ configuration) (Fig.~\ref{Fig2} (h)) whereas the unoccupied P-p states appear 7 eV above E$_F$ (Fig.~\ref{Fig2} (d)). 

To further analyze the degree of $pd$-hybridization, hopping integrals ($\vert t_{pd} \vert$) between O-p states and Cu-d states have been calculated using MLWF. 
To construct a basis of MLWF we employ a wide energy window that includes the full Cu- d manifold, O-p, Cl-p, and Cs-p states. The agreement between the band structure
obtained from Wannier function interpolation and that derived
from the DFT calculation is excellent (see Appendix Fig.~\ref{Fig6}). The spatial spreads of the Wannier functions are small ( $\sim$ 1 \AA$^2$). 
The derived dominant $\vert t_{pd} \vert$ integrals are listed in Table ~\ref{Tab2}. Overall, the $\vert t_{pd} \vert$ hoppings increase in the P-system due to the induced chemical pressure effect upon volume reduction, in agreement with the above described increase in bandwidth in the P-material. Only the $\vert t_{pd} \vert$ between Cu$_k$-d$_{x^2-y^2}$  and O$_k$-p$_{x,y}$  is unaffected (0.55 eV), whereas the  hopping integrals between Cu$_h$-d$_{z^2}$ and O$_{k}$-p$_{x,y}$ as well as between Cu$_k$-d$_{x^2-y^2}$ and O$_{h}$-p$_{x,y}$ increase significantly in P-averievite (0.97 vs 0.83 eV and 1.15 vs 1.04 eV). 

\subsection{Magnetic properties}
\begin{figure}
\vspace{-0.3cm}
\begin{center}
\includegraphics[width=\columnwidth]{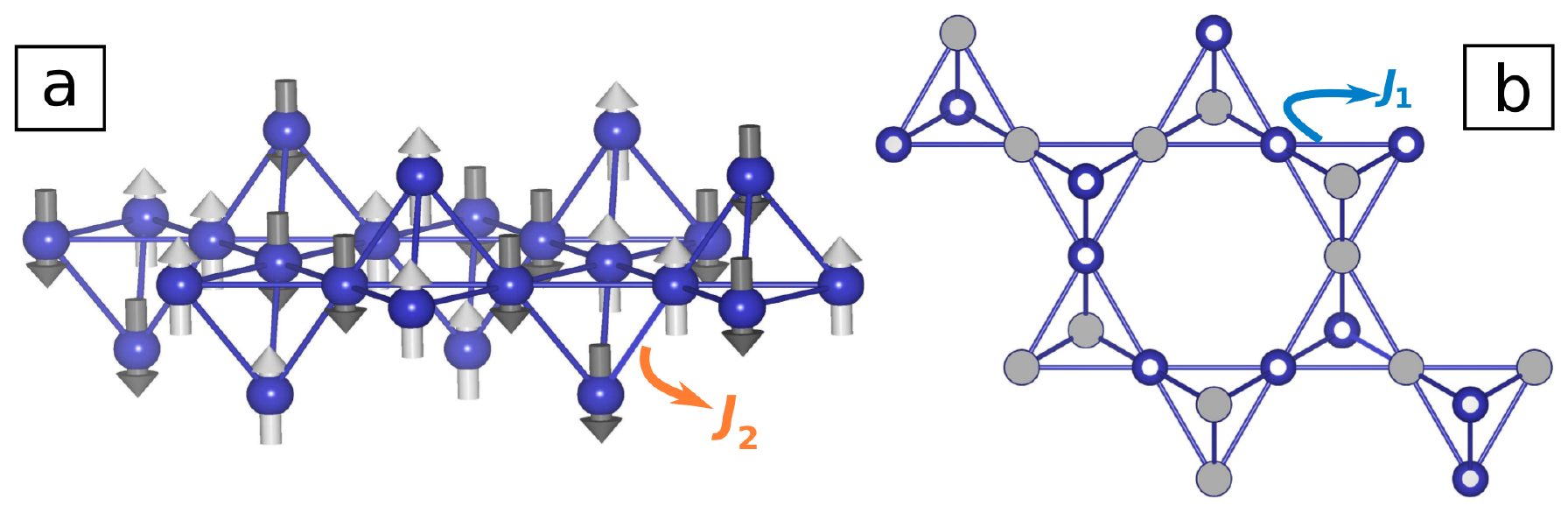}
\caption{\label{Fig3}(Color online) (a) Magnetic ground state for P- and V-averievite. The two dominant AFM  exchange interactions between kagome-Cu (J$_{1}$) (b) and kagome-to-honeycomb Cu atoms (J$_{2}$) (a) are shown.}
\end{center}
\end{figure}

\begin{figure}
\includegraphics[width=8.5cm]{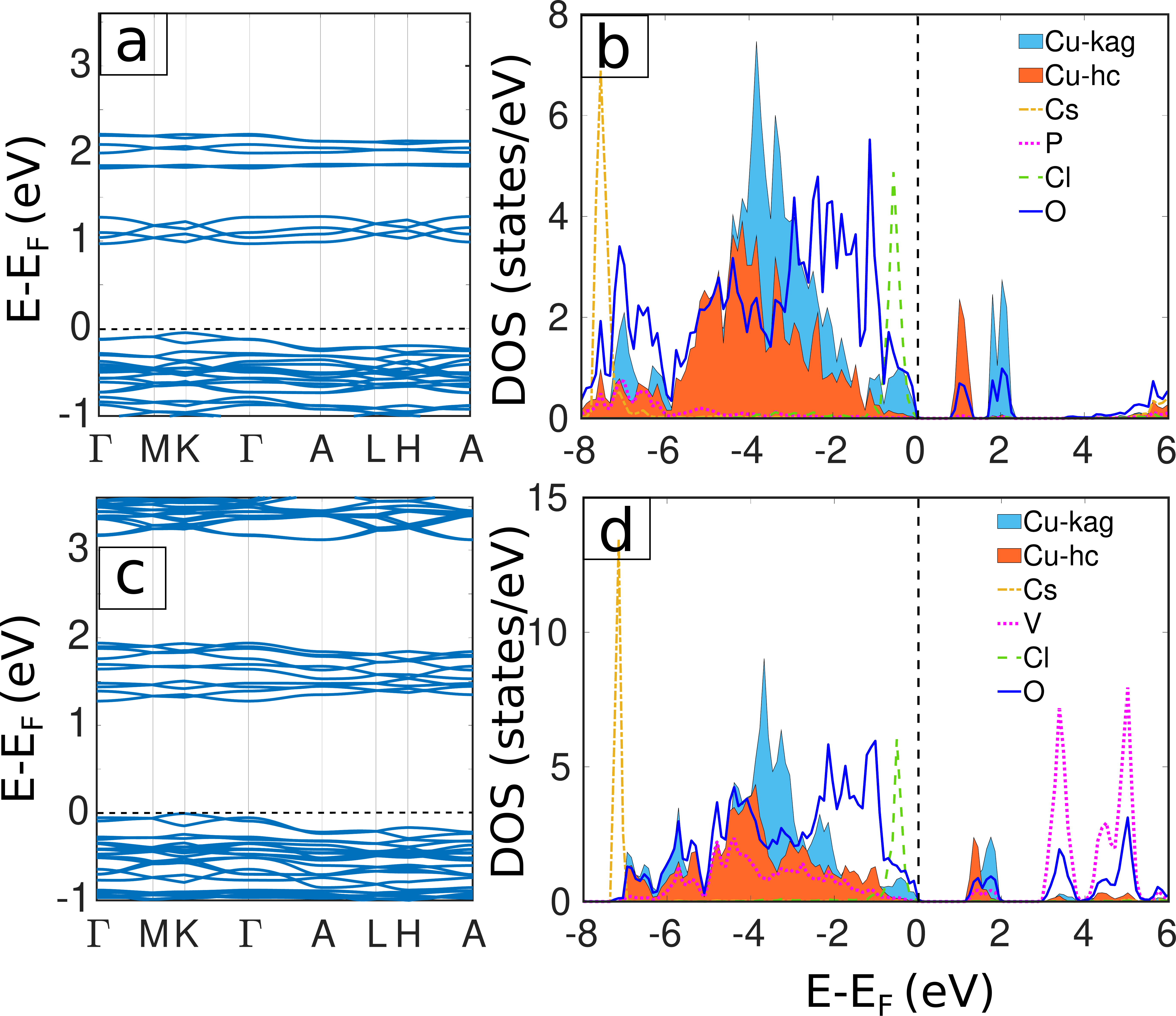}
\caption{\label{Fig4}(Color online) (a, c) GGA+U band structure, and (b, d) atom projected density of states  of P-averievite (top-panels) and V-averievite (bottom panels) in the AFM ground state shown in Fig. 3.}
\end{figure}

 GGA+U calculations in different  magnetic configurations have been performed  in a 2$\times$2$\times$1 supercell containing twenty inequivalent Cu sites. The magnetic ground state for V and P-averievite is depicted in Fig.~\ref{Fig3}. This configuration corresponds to AFM nearest-neighbor (NN) coupling in the kagome planes and between the kagome and honeycomb coppers, which gives rise to an AFM configuration on each Cu-tetrahedron. Within GGA+U,  both compounds exhibit an insulating ground state in this magnetic configuration. The corresponding DOS and band structure are shown in Fig.~\ref{Fig4}. In P-averievite, the band gap is $\sim$0.9 eV, smaller than that for V-averievite $\sim$1.3 eV. 
Right below the Fermi level, the O-p and Cu-d character described in the non-magnetic calculations is kept, but there is also a large contribution of Cl-p states to the DOS. This is due to the orientation of the CuO$_4$ units in the kagome planes, with the Cu-d$_{z^2}$ orbitals pointing towards the Cl ion, giving rise to a high degree of hybridization. The unoccupied honeycomb and kagome Cu-d  states are separated in energy in the phosphate compound, whereas they overlap in the vanadium material. 
The magnetic moment of Cu atoms in both systems is 0.7 $\mu_B$, consistent with a d$^9$ configuration (S = 1/2). No sizable moments are developed on other atoms. 

The exchange-coupling constants for P- and V-averievite can
be obtained by mapping the GGA+U energy differences for
different magnetic configurations (see Appendix~\ref{appB} for details) to a spin-1/2 Heisenberg
Hamiltonian of the form:

\begin{equation} 
H= \sum\limits_{i< j} J_{ij} S_i \cdot S_j 
\end{equation}

The leading terms are the two nearest-neighbor AFM couplings mentioned above: J$_1$ between Cu ions in the kagome plane and J$_2$ between kagome and
honeycomb Cu ions (see Fig. \ref{Fig3}). Further exchange terms are significantly weaker. Specifically, we find J$_{1}$= 235 K and J$_{2}$= 284 K for P-averievite whereas for V-averievite we obtain J$_{1}$=228 K and J$_{2}$=52 K. Notably, J$_{2}$ is five times larger in P-averievite. The increase in J$_2$ in the phosphate material is consistent with the above described hopping integrals and with the increase in Cu$_k$-O$_k$-Cu$_h$ bond angle with respect to its V-analog.
 
 In a pyrochlore lattice (or slab) formed by magnetic ions, there is a high degree of frustration when the nearest-neighbor interactions are AFM. This is effectively the situation in P- and V- averievite in which we find two AFM interactions within a given Cu tetrahedron (J$_1$ and J$_2$). In this situation, the more comparable these two AFM interactions are, the larger the degree of magnetic frustration  ~\cite{Rami_Tc, Pyro_slab_MC}. This is effectively what seems to happen in P-averievite where we find comparable J$_1$ and J$_2$ and for which the experimentally determined frustration index is larger than that of its V-counterpart (f$_{V-averievite}$= $\Theta_{CW}$/T$_t$ $\sim$ 8~\cite{BotanaPRB}, f$_{P-averievite}$ $\sim$ 13~\cite{WiniACS} where $\Theta_{CW}$ is the Curie-Weiss temperature, and T$_t$ the temperature of the magnetic transition).  We note that spin-glass-like transitions at low temperature have been reported in other pyrochlore-slab materials with strong AFM interactions (i.e. SrCrGaO compounds)~\cite{Pyro_slab, Rami_Tc, Pyro_slab_MC} This seems to be consistent with experimental data in P-averievite that show a lowering of the magnetic transition temperature with respect to the V-based material and point out to the possibility of a spin-glass transition in the P-system \cite{WiniACS}.


\subsection{Effects of Zn-doping}
As pointed out for V-averievite \cite{BotanaPRB}, substitution of Cu$^{2+}$ by non-magnetic Zn$^{2+}$ in the honeycomb layers is an interesting strategy to suppress the inter-layer coupling and long-range magnetic order.  We have performed this substitution for P-averievite, obtaining (CsCl)Cu$_3$Zn$_2$P$_2$O$_{10}$ (Zn-substituted P-averievite). Substitution of Zn ions in the honeycomb plane is energetically more favorable than substitution in the kagome layer due to the differing oxygen environments in the two layers. In this situation, the kagome planes, separated by a large distance (8.45 \AA), are the only magnetically active ones.

\begin{figure}
\vspace{0.1cm}
\includegraphics[width=8.5cm]{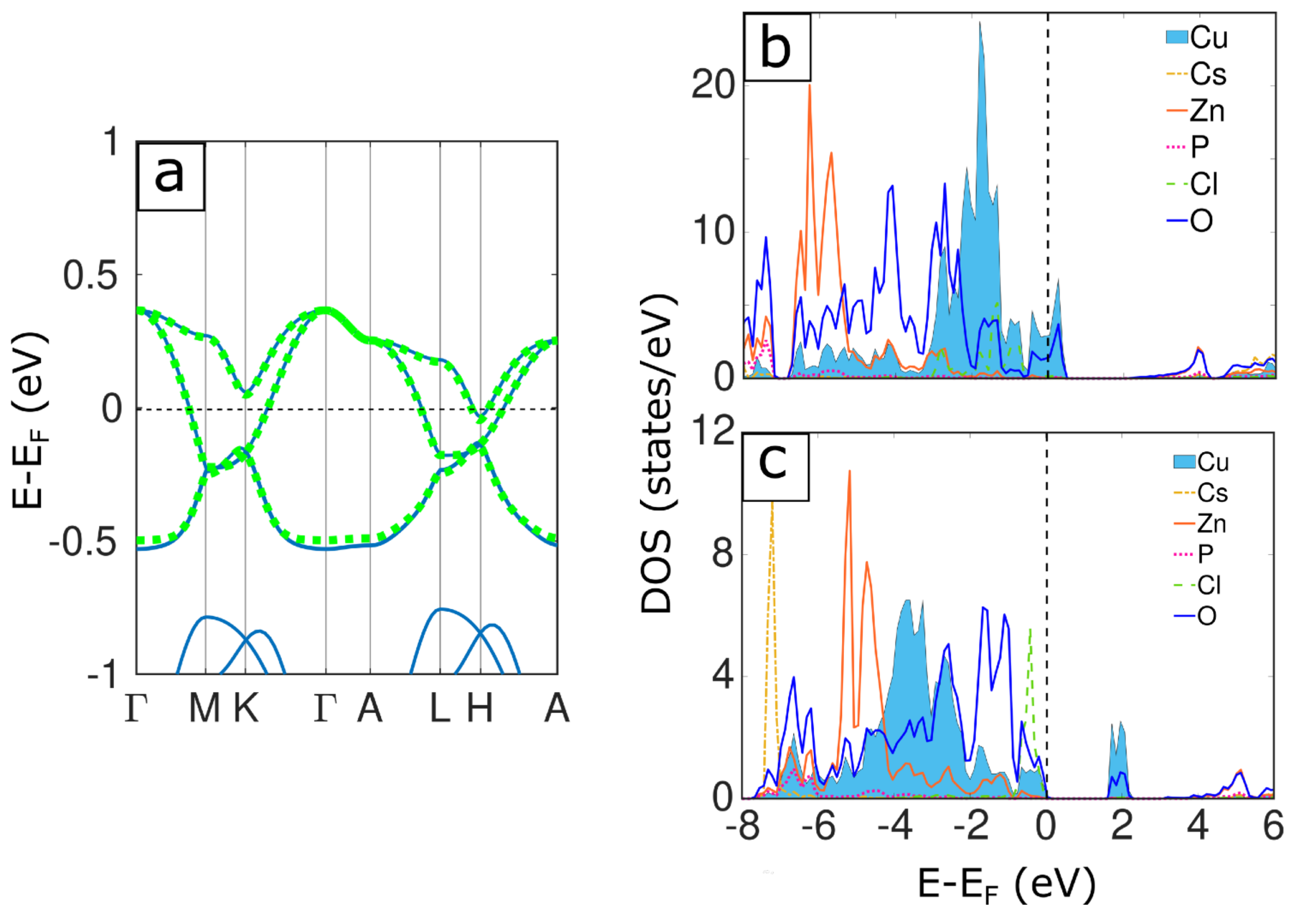}
\caption{\label{Fig5} (Color online) DOS and band structure of Zn-substituted P-averievite.  (a) Non-magnetic GGA band structure (solid blue line) and the corresponding Wannier fitting to the three Cu-d$_{x^2-y^2}$ bands (green dotted line). (b) Non-magnetic GGA atom-projected DOS and (c) Atom-projected DOS as obtained within GGA+U for an AFM spin configuration within kagome planes.}
\end{figure}

Fig.~\ref{Fig5} shows the corresponding GGA (non-magnetic) and GGA+U (AFM) electronic structure of (CsCl)Cu$_3$Zn$_2$P$_2$O$_{10}$. In the non-magnetic GGA band structure (Fig.~\ref{Fig5}(a)), there are  three isolated bands around the Fermi level of kagome Cu-d$_{x^2-y^2}$ character only, due to the removal of the honeycomb coppers. There are clear Dirac crossings at K and H points akin to the kagome single-orbital tight-binding model \cite{kagome_TB}. The same type of behavior has been reported in Zn-substituted V-averievite \cite{BotanaPRB}. We have extracted tight-binding parameters for Zn-substituted averievite using MLWF. To  construct a basis of MLWF in this case we use a narrow energy window of $\sim$ 0.9 eV (see Fig.~\ref{Fig5} (a)) including only the Cu-d$_{x^2-y^2}$ states. The corresponding fit (green dotted lines in Fig.~\ref{Fig5}(a)) matches well the band dispersion obtained from DFT calculations indicating a faithful transformation to MLWF. The spread of the derived Wannier functions is small ($\sim$ 1 \AA$^2$).

The nearest-neighbor hopping integral is -123 meV, and the next nearest-neighbor hopping integral is -28 meV. These values are comparable to those of the hopping integrals of Zn-substituted V-averievite \cite{BotanaPRB}.  The non-magnetic GGA DOS (Fig.~\ref{Fig5}(b)) reveals that the Cu-d bands crossing E$_F$ hybridize strongly with O-p states, and that the valence bandwidth ($\sim$8 eV) is reduced with respect to that of the parent compound. Zn$^{2+}$-3d states are completely filled, and lie at lower energies. 

The GGA+U DOS corresponding to the lowest energy AFM spin configuration within kagome planes is depicted in Fig.~\ref{Fig5}(c). In the presence of U, the system is insulating with an estimated bandgap of 1.7 eV, increased by 0.8 eV with respect to the parent compound. Like in P-averievite, the GGA+U DOS has main contributions from Cu-d, O-p, and Cl-p states right below E$_F$. Zn-d states appear at lower energies. Unoccupied states are mostly Cu-d in character, keeping a high degree of hybridization with O-p states. The magnetic moments of Cu atoms in the kagome plane remain the same as in the parent material $\sim$ 0.7 $\mu_B$. The resulting exchange interactions for Zn-substituted averievite can be derived using the methodology described above. A dominant AFM J$_1$ = 145 K within the kagome plane is obtained from our calculations, 90 K lower than that for undoped P-averievite. 
A decrease in J$_1$ upon Zn-substitution has also been reported in V-averievite \cite{BotanaPRB}, but the value of J$_1\sim$ 170 K is higher in that case. Experiments will have to be performed to confirm if substitution of Cu$^{2+}$ by non-magnetic Zn$^{2+}$ can suppress long-range magnetic order in P-averievite as well.

\section{Conclusions}
\label{conc}

We have analyzed the effects of V by P substitution in the electronic and magnetic properties of the frustrated quantum magnet averievite. Both P- and V- averievite contain Cu$^{2+}$ kagome layers sandwiched between  Cu$^{2+}$-P$^{5+}$/Cu$^{2+}$-V$^{5+}$ honeycomb planes with pyrochlore slabs made of corner-sharing Cu-tetrahedra being formed. The tetrahedral geometry introduces magnetic  frustration, as interactions within the Cu$_4$ units are AFM. Structural changes arise due to chemical pressure as the ionic radius of P$^{5+}$ in tetrahedral coordination is over two times smaller than that of V$^{5+}$. Our calculations reveal that the nearest-neighbor AFM coupling (J$_1$) between kagome Cu atoms remains the same $\sim$235 K in P- and V-averievite. In contrast, the inter-layer AFM coupling (J$_2$) between kagome and honeycomb Cu ions is 284 K in P-averievite, five times larger than the value of J$_2$ in its V-counterpart. The stronger J$_2$ increases the degree of magnetic frustration within Cu-tetrahedra in the phosphate material and is in agreement with the increase in the experimentally reported frustration index for P-averievite.  Further, long-range magnetic order could be suppressed in Zn-substituted P-averievite, as the inter-layer coupling is absent and the kagome spin-1/2 planes are the only magnetically active ones, making it a good candidate for QSL behavior.

As averievite is an oxide (vs. traditional hydroxide  platforms for spin-1/2 kagome physics) we anticipate this material should pose some advantages: 1) it should be less prone to disorder as Zn should substitute on the
honeycomb copper sites, based on crystal chemical considerations. 2) it shows a larger degree of p-d hybridization and should hence be more likely to promote metallicity (and possibly superconductivity). Based on these considerations, we hope our calculations stimulate further experiments in Zn-doped averievite and in other oxide spin-1/2 kagome systems.

\section{Acknowledgments}
ASB acknowledges NSF-DMR grant 1904716. DD acknowledges ASU for startup funds.  We acknowledge the ASU Research Computing Center for HPC resources.

\appendix 
\section{Wannier fitting of the DFT bands of undoped P- and V-averievite}
\label{appA}
\begin{figure}[ht]
\vspace{-0.8cm}
\includegraphics[width=8.6cm]{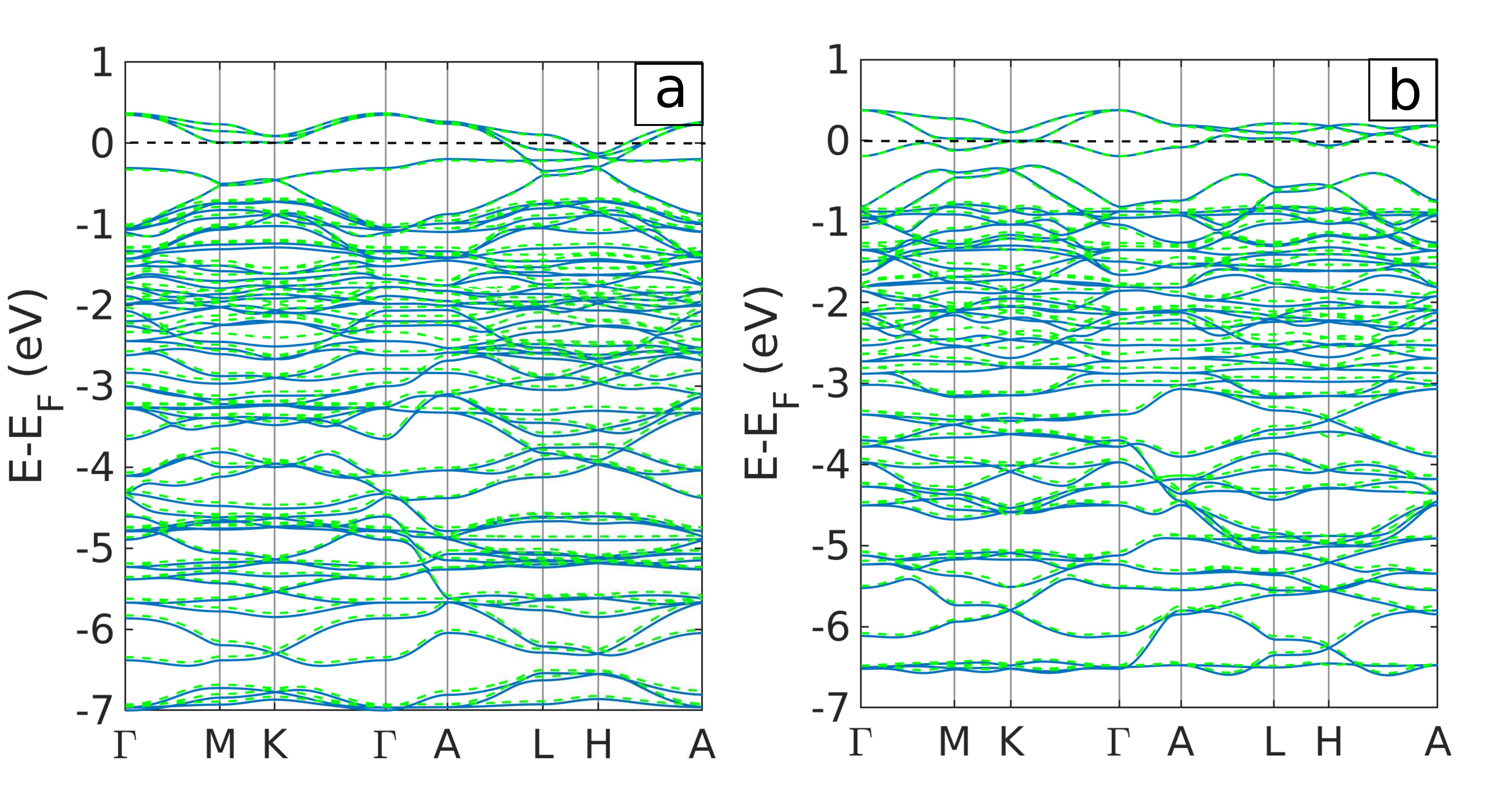}
\caption{\label{Fig6} (Color online) Non-magnetic GGA band structure (solid blue line) and the corresponding Wannier fitting (green dotted line) for (a) V-averievite and (b) P-averievite.}
\end{figure}

\section{Calculation of the exchange couplings}
\label{appB}
\begin{figure}
\vspace{-0.8cm}
\includegraphics[width=8.8cm]{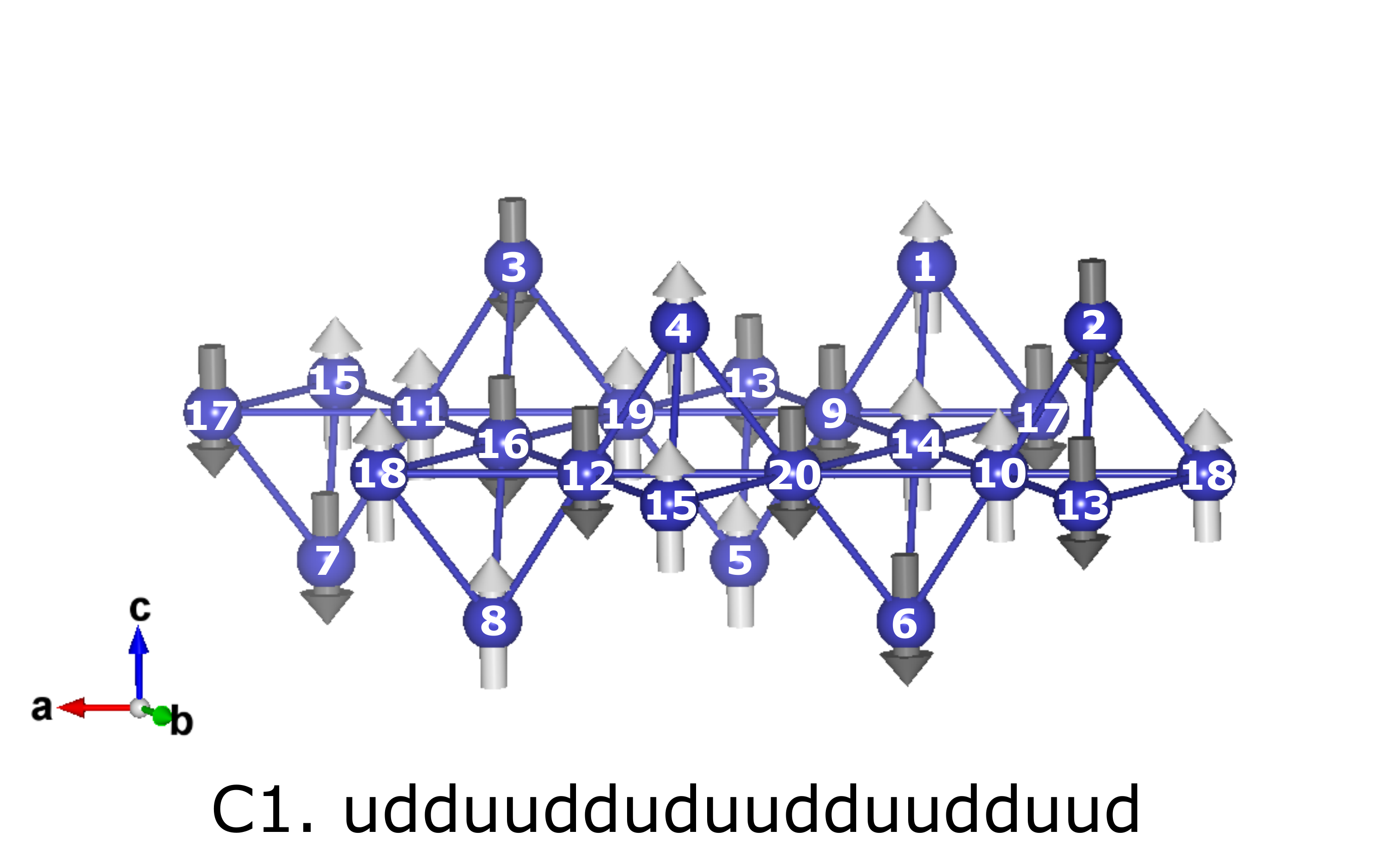}
\caption{\label{Fig7} (Color online) C1 represents the magnetic ground state defined bas udduudduduudduudduud, where u$\equiv \uparrow$ and d$\equiv \downarrow$, stand for majority and minority spins, respectively.}
\end{figure}

\begin{figure}[ht]
\includegraphics[width=8.6cm]{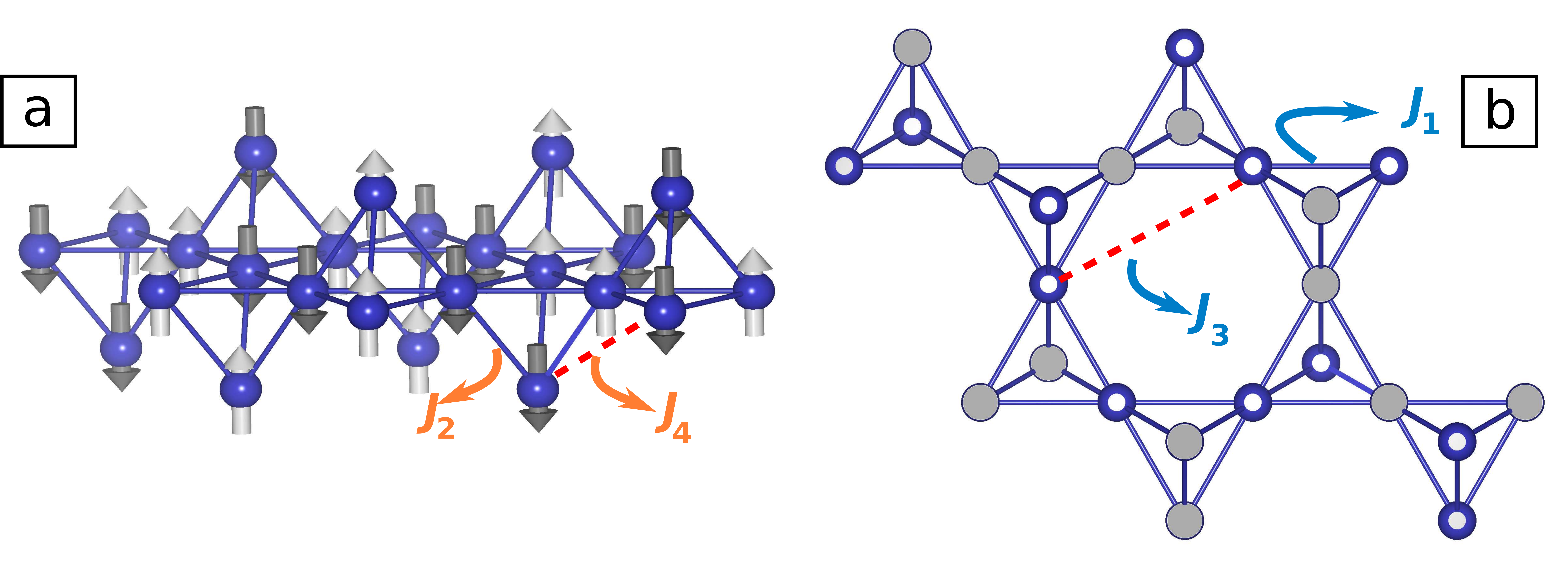}
\caption{\label{Fig8} (Color online) Magnetic ground state for P- and V-averievite. Nearest-neighbor (NN) and next-nearest-neighbor (NNN) exchange paths are shown (a) between kagome and honeycomb Cu atoms (J$_2$ and J$_4$) and (b) between kagome-Cu (J$_1$ and J$_3$). Red dotted line represents the NNN exchange path.
}
\end{figure}

 We have estimated magnetic exchange couplings up to next-nearest-neighbors (NNN) fitting the energy differences of different magnetic configurations to a Heisenberg model. Ten spin configurations were constructed in a 2$\times$2$\times$1 supercell with 20 Cu atoms: C1 (ground state)  udduudduduudduudduud (the labeling is defined in Fig.~\ref{Fig7}), C2. udduudduduuddddudddd, C3. ddduudduduudddduduud, C4. udduudduddddduudduud, C5. udduudduduuduuuuduud,  C6. udduuddududdduudduud, C7. udduudduduudduuuduud, C8. udduudduduudddduduud, C9. udduuddududdddduduud, and C10. uuuuuuuuuuuuuuuuuuuu. The DFT energy differences ($\Delta E$) between each of these configurations and the magnetic ground state (C1) are listed in Table~\ref{Tab3}. Fitting the DFT energies (of five different spin configurations at a time) to a spin-1/2 Heisenberg model we obtain, with small errors in the determination, the four exchange couplings (J$_{1-4}$) whose paths shown in Fig.~\ref{Fig8}. For P-averievite (V-averievite), J$_1$ (in-plane NN) = 234.9 (227.8) K, J$_2$ (out-of-plane NN) = 284.3 (51.7) K, J$_3$ (in-plane NNN) = -5.4 (-3.2) K, and J$_4$ (out-of-plane NNN) = 0.5 (1.7) K. It is evident from our calculations that J$_1$ and J$_2$ are the leading terms, whereas J$_n$ for further exchange paths are significantly weaker.

\begin{table}
\centering
\begin{tabular}{p{2.5cm} p{2.5cm} p{2.5cm} p{0.01cm}}
\hline
 \hline
\centering $\Delta E$ (meV)/f.u. & \centering (CsCl)Cu$_5$P$_2$O$_{10}$ & \centering (CsCl)Cu$_5$V$_2$O$_{10}$ & \\
\hline
\centering C1 & \centering 0.0 & \centering 0.0 & \\
\centering C2 & \centering 15.2 & \centering 14.8 &\\ 
\centering C3 & \centering 24.5 & \centering 16.6 &\\ 
\centering C4 & \centering 12.2 & \centering 2.1 &\\
\centering C5 & \centering 10.2 & \centering 9.9 &\\
\centering C6 & \centering  6.1 & \centering 1.1 &\\
\centering C7 & \centering  5.1 & \centering 4.8 &\\ 
\centering C8 & \centering 15.3 & \centering 15.1 &\\ 
\centering C9 & \centering 22.9 & \centering  16.6 &\\
\centering C10 & \centering 111.1 & \centering 63.4 &\\
\hline
\hline
\end{tabular}
\caption{Energy difference per formula unit (f.u.) between different magnetic configurations with respect to the magnetic ground state (C1) defined by  $\Delta E$=0 .}
\label{Tab3}
\end{table}

%

\end{document}